\documentclass{article}
\usepackage{spconf,amsmath,graphicx}
\usepackage{adjustbox}
\usepackage{xcolor}
\usepackage{multirow}
\usepackage{booktabs}
\usepackage{url}
\usepackage[colorinlistoftodos]{todonotes}
\usepackage{amssymb}

\title{Corpus Synthesis for Zero-shot ASR domain Adaptation using \\ Large Language Models}
\name{
\begin{tabular}{c} Hsuan Su$^\heartsuit$$^\diamondsuit$\sthanks{Work done when interning at Apple.} \qquad Ting-Yao Hu$^\diamondsuit$ \qquad Hema Swetha Koppula$^\diamondsuit$ \qquad Raviteja Vemulapalli$^\diamondsuit$ \\
Jen-Hao Rick Chang$^\diamondsuit$ \qquad Karren Yang$^\diamondsuit$ \qquad Gautam Varma Mantena$^\diamondsuit$ \qquad Oncel Tuzel$^\diamondsuit$\end{tabular}
\vspace{-0.15in}}
\address{$^{\heartsuit}$National Taiwan University \qquad $^\diamondsuit$Apple}
\begin{document}
%\ninept
%
\maketitle

\begin{abstract}
While Automatic Speech Recognition (ASR) systems are widely used in many real-world applications, they often do not generalize well to new domains and need to be finetuned on data from these domains. 
% However, data from target domains may not be readily available in many scenarios. 
However, target-domain data usually are not readily available in many scenarios.
In this paper, we propose a new strategy for adapting ASR models to new target domains without any text or speech from those domains.
% In this paper, we propose a new paradigm that adapts ASR models to new target domains (characterized by changes in text corpus) without using any text or speech from these domains. 
To accomplish this, we propose a novel data synthesis pipeline that uses a Large Language Model (LLM) to generate a target domain text corpus, and a state-of-the-art controllable speech synthesis model to generate the corresponding speech. 
We propose a simple yet effective in-context instruction finetuning strategy to increase the effectiveness of LLM in generating text corpora for new domains.
% We also use instruction finetuning and in-context learning to increase the effectiveness of the LLM in generating text corpora for new domains.
Experiments on the SLURP dataset show that the proposed method achieves an average relative word error rate improvement of $28\%$ on unseen target domains without any performance drop in source domains.

\end{abstract}


\section{Introduction}
\label{sec:intro}

Adapting an End-to-End (E2E) Automatic Speech Recognition (ASR) system to new target domains is a challenging task due to the limited availability of paired speech-text data.
Recently, text-only adaptation methods %\cite{meng2021internal,pylkkonen2021fast,deng2023adaptable,sato2022text,mittal2022situ,joshi2022simple,bataev2023text} 
\cite{meng2021internal,pylkkonen2021fast,deng2023adaptable,mittal2022situ,joshi2022simple,bataev2023text}
have been developed to address the data scarcity problem.
Some of these works~\cite{joshi2022simple,bataev2023text} use a Controllable Speech Synthesis (CSS) model to generate speech for the target domain text corpus, and create a paired  dataset with real text and synthetic speech for ASR model adaptation. However, in many scenarios, collecting a target domain text corpus may be costly, time consuming, or even infeasible due to privacy concerns.
%
%For example, when a virtual assistant with a speech interface is updated for a new feature, there is no real user text data available before the launch of this feature.
\begin{figure}[htp!]
    \begin{adjustbox}{width=\linewidth}
    \centering
    \includegraphics{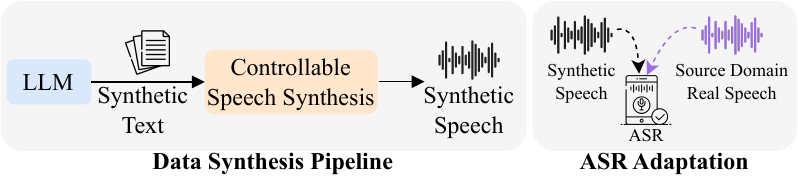}
    \end{adjustbox}
    \caption{\textbf{Data Synthesis Pipeline.} The pipeline consists of a LLM and a CSS model. We use a LLM to generate text corpus, and then synthesize speech data using a CSS model. \textbf{ASR Adaptation} - The synthetic target domain data is used to finetune a pretrained ASR model along with real speech from source domains. See Section~\ref{sec:method} for details.}
     \label{fig:framework}
\end{figure}
% \begin{figure}[htp!]
%     \begin{adjustbox}{width=\linewidth}
%     \centering
%     \includegraphics{pics/new_example.pdf}
%     \end{adjustbox}
%     \caption{\textbf{Examples of instruction finetuning and zero-shot inferencing LLMs.} We explicitly relate domain name $d$ to text $x$w by reformulating an instruction to finetune LLMs. During the inference time, we use an unseen target domain with instruction to inference LLMs. See \ref{sec:instuct} for details.}
%     \label{fig:example}
% \end{figure}

%[Why large language model?]
Large Language Models (LLMs) have recently been shown to work extremely well on numerous natural language processing tasks, especially in few/zero-shot settings.
This motivates us to leverage LLMs for adapting ASR models to new domains without any speech or text data from those domains.
%
%\todo{Add the citation for `previous works` here.}
%
Previous works exploit LLMs in ASR systems during inference using re-scoring \cite{Dingliwa2022,ma2023can} or fusion \cite{li2023prompting} techniques, and they suffer from the costly overhead of LLM inference.  
In contrast, we use LLMs to generate target domain synthetic data for adapting a pretrained ASR model (see Fig.~\ref{fig:framework}).

%First, we generate a synthetic text corpus for the target domain by prompting an LLM. 
%To improve the quality of the synthetic text corpus, we propose a simple yet effective in-context instruction finetuning strategy for LLMs using text from related source domains.
%To improve the quality of the synthetic text corpus, we apply instruction finetuning (IF) \cite{chung2022scaling} and in-context learning (ICL) \cite{brown2020language} to guide the generation process of LLMs.
%
First, we generate a synthetic text corpus by prompting an LLM with the target domain name (a word or short phrase).
To improve the quality of the synthetic text corpus, we propose a simple yet effective in-context instruction finetuning (ICIF) strategy.
Assuming that the pretrained ASR model has been trained on a source dataset with multiple domains (e.g. a personal assistant with a set of existing features),
ICIF learns to relate domain names to the knowledge of LLMs from source text sentences. 
%It uses a set of source domain text sentences with their domain names annotated, and relates the domain names to the knowledge of LLMs (see Section~\ref{sec:instuct}).
Then, we use a state-of-the-art CSS model~\cite{chang2022style} to generate speech corresponding to the synthetic texts.
Finally, the fully synthetic paired speech-text corpus is used to finetune a pretrained ASR model, improving performance on the target domain of interest while retaining the performance on the source domain.

\textbf{Major contributions}: 
(1) We demonstrate that text corpus synthesis using LLMs enables zero-shot ASR domain adaptation. % of ASR models.
%
%(2) We propose a simple and effective instruction finetuning strategy for LLMs to further improve the quality of synthetic text corpus.
(2) Our proposed in-context instruction finetuning strategy improves the quality of the synthetic text corpus resulting in significant gains in the final ASR performance.
(3) We show that the proposed data synthesis pipeline achieves an average of $28 \%$ relative Word Error Rate (WER) reduction on unseen target domains in the SLURP dataset~\cite{bastianellietal2020slurp}.
%\todo{Add citation} 
%

% \begin{figure}[htp!]
%     \begin{adjustbox}{width=\linewidth}
%     \centering
%     \includegraphics{pics/new_framework.pdf}
%     \end{adjustbox}
%     \caption{\textbf{Corpus Synthesis Pipeline.} The pipeline consists of a LLM and an CSS model. We first prompt the pretrained or instruction finetuned LLM to synthesize text corpus, and prepare speech data with decent CSS model. See Section \ref{sec:method} for details.}
%      \label{fig:framework}
% \end{figure}
\section{Related Works}

%\subsection{Text only ASR adaptation}

%\subsection{Data generation with LLMs}
%Many works have shown LLMs have ability to synthetsize data and can be used for downstream tasks. 
Many works have shown that LLMs can synthesize data useful for downstream tasks. 
Ye et al.~\cite{ye-etal-2022-zerogen}, Yoo et al.~\cite{yoo-etal-2021-gpt3mix-leveraging}, and Meng et al.~\cite{meng2022generating} prompt LLMs with handcrafted instructions to generate data for finetuning downstream models. 
%We further use the source domain data to instruct finetune LLMs beforehand to improve the format consistency for generation text.
In this work, we finetune the LLMs with instruction data to improve the format consistency of the generated text corpus.

%\subsection{LLM for ASR}
Some previous works also use LLMs to improve the performance of ASR. 
Dingliwa et al.~\cite{Dingliwa2022} and Ma et al.~\cite{ma2023can} conduct second-pass re-scoring using the perplexity score from LLMs. 
Li et al.~\cite{li2023prompting} propose deep LLM-fusion, which integrates an LLM into the decoder of an encoder-decoder based E2E ASR model.
While these methods improve performance, they require LLM inference during ASR decoding, which increases the computational cost.
In contrast, our method transfers knowledge from the LLM to an ASR model through a synthetic text corpus.
%\subsection{Synthetic speech for ASR training}

%\vspace{-0.1in}
\section{Methodology}
\label{sec:method}
%\vspace{-0.1in}
Fig.~\ref{fig:framework} shows an overview of the proposed approach.
Our pipeline consists of a LLM, a CSS model, and a pretrained ASR model.
Given a target domain of interest $d_t$, we generate a fully synthetic text-speech paired corpus $C_t=\{(x^t_i, y^t_i)\}_{i=1}^N$, where $x^t_i$, $y^t_i$ are the text content and speech signal of the $i$-th sample respectively.
To do this, we first generate a text sentence $x^t_i \sim p_{LLM}(x | d_t)$ from the LLM conditioned on $d_t$. Then, we synthesize the corresponding speech $y^t_i \sim p_{CSS}(y|x_i^t)$ using the CSS model.
Finally, we use the synthetic text-speech data to finetune the ASR model for target domain $d_t$. %, and the finetuned ASR model obtains better performance on the testing set of target domain.
%
% Although preliminary experiment shows that naive usage of this pipeline already provides decent improvement, we find that it is important to address the discrepancy between real and synthetic text data distribution.
% %
% Considering that the LLM is trained on a very large-scale, general text corpus, it is difficult to produce high quality in-domain data with a target domain name $d_t$.
% %
% Thus, we further utilize a text corpus $C_s={(x^s_j, d^s_j)}$ derived from the training set of the pretrained ASR, where $d^s_j$ is the domain name of the $j-th$ sample.  
% %
% To this end, we propose a simple yet effective instruction finetuning strategy for LLM with $C_s$, and observe that it improves the quality of the synthetic text corpus.

\vspace{-0.05in}
\subsection{Text Synthesis with LLMs}
\vspace{-0.05in}
    % LLaMA 
Our goal is to synthesize a text corpus that matches the text distribution of a given target domain $d_t$. To this end, we ask LLMs which are pretrained on trillions of text tokens to generate sentences relevant to the target domain using the prompt: ``\textit{Please generate a sentence related to $d_t$:}''.
Our initial experiments show that naively prompting off-the-shelf LLMs in our pipeline leads to some ASR improvement on the target domain. However, the quality of the generated text and its relevance to the target domain are both insufficient. % high.
%it is important to address the discrepancy between real and synthetic text data distribution.
Since off-the-shelf LLMs are trained on large-scale general text corpora, it is difficult for them to produce high quality in-domain text using only the target domain name $d_t$.
% 
%To address this issue, we propose a simple yet effective instruction finetuning strategy that improves the ability of LLMs to generate in-domain data when prompted with the target domain name. 
%To address this issue, we adapt in-context instruction finetuning \cite{ye2023incontext} to guide the decoding process of LLMs, improving their ability to generate in-domain like data when prompted with the target domain name.
To address this issue, we propose a simple yet effective in-context instruction finetuning (ICIF) strategy that improves the ability of LLMs to generate in-domain text when prompted with the target domain name.

\vspace{-0.05in}
\subsubsection{In-Context Instruction Finetuning (ICIF)}
\label{sec:instuct}
%\textbf{Instruction Finetuning (IF):}
\vspace{-0.05in}

Our proposed in-context instruction finetuning (ICIF) strategy combines instruction finetuning (IF) \cite{chung2022scaling} with demonstration or in-context learning (ICL) \cite{brown2020language} using a unified instruction format.
Specifically, we first finetune the LLM with instruction data. Then, during inference, we prompt the LLM with additional demonstrations in the same instruction format.

\begin{figure}[tp!]
    \begin{adjustbox}{width=\linewidth}
    \centering
    \includegraphics{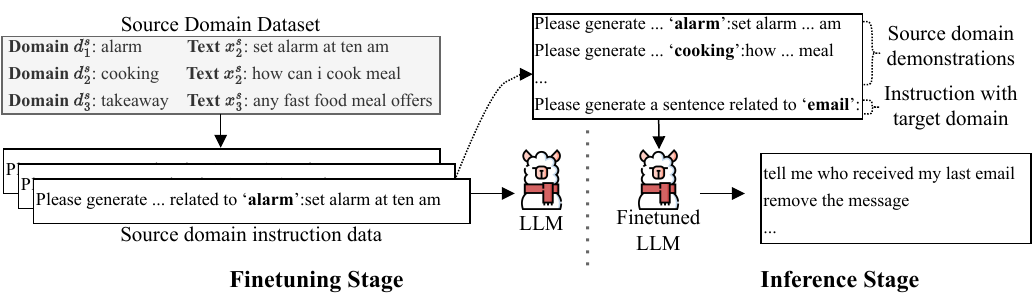}
    \end{adjustbox}
    \caption{\textbf{Illustration of In-Context Instruction Finetuning.} We explicitly relate domain name $d$ to text $x$ by reformulating an instruction to finetune the LLMs. During inference, we use source domain demonstrations and an unseen target domain (\textit{email} in the figure) instruction to prompt the LLMs. }
    \label{fig:example}
\end{figure}

To construct the instruction data and demonstrations, we use a source text corpus $C_s=\{(x^s_j, d^s_j)\}_{j=1}^M$, which contains text $x^s_j$ from source domains $d^s_j$ distinct from $d_t$.
%
%we use a text corpus $C_s=\{(x^s_j, d^s_j)\}$ derived from the training set of the pretrained ASR, where $d^s_j$ is the domain name of the $j$-th sample.
%
As shown in Fig.~\ref{fig:example}, we reformulate each $(x^s_j, d^s_j) \in C_s$ as a natural language instruction-- ``\textit{Please generate a sentence related to $d^s_j$~:~$x^s_j$}''-- and finetune the LLM on these instructions.
In the inference stage, we prepend a subset of these instructions from the source domain to the original prompt for the unseen target domain (``\textit{Please generate a sentence related to $d_t$:}'') as additional demonstrations. The LLM uses the extended prompt to generate a text sentence in target domain. %In practice, the source text corpus can be derived from the source domain of the pretrained ASR model.

Our ICIF strategy learns the structure and format of the source corpus $C_s$, and relates the target domain name $d_t$ to the knowledge from pretrained LLMs.
As shown in Section~\ref{sec:size}, the resulting synthetic text corpus comprises high quality, diverse sentences which are semantically related to the unseen target domain.
 % LLaMA \cite{touvron2023llama} is a state of the art foundation model that pretrained on 1.4T tokens with transformer-based architecture \cite{vaswani2017attention}. LLaMA-13B have shown to be able to outperforms GPT-3 (175B) \cite{brown2020language} on most benchmarks, and LLaMA- 65B is competitive with the best models, Chinchilla-70B \cite{hoffmann2022training} and PaLM-540B \cite{chowdhery2022palm}. In our experiments, we used LLaMA-7B \cite{touvron2023llama} to synthesize text data.
\begin{table*}[t]
\begin{adjustbox}{width=\linewidth}
\begin{tabular}{cccccccccccccccccccc}
\toprule
Zero-shot (WER)                                                   & \multicolumn{18}{c}{Target Domains}                                                                                                                                                                                                                                                                                                                & \multirow{2}{*}{Average} \\ \cmidrule{1-19}
Methods                                                           & Alarm            & Audio            & Calendar         & Cooking          & Datetime         & Email            & General          & IOT              & Lists            & Music            & News             & Play             & QA               & Recommendation   & Social           & Takeaway         & Transport        & Weather          &                          \\ \midrule
\begin{tabular}[c]{@{}c@{}}Source domain ASR\\ (Baseline)\end{tabular} & 8.0              & 13.1             & 12.8             & 18.2             & 11.2             & 19.0             & 14.4             & 19.2             & 14.6             & 10.5             & 15.3             & 24.8             & 22.3             & 15.7             & 26.3             & 26.5             & 17.1             & 12.9             & 16.77                    \\ \midrule
ICIF                                                          & \textbf{4.90}    & \textbf{7.50}    & \textbf{10.27}   & \textbf{9.93}    & \textbf{8.33}    & \textbf{12.70}   & \textbf{13.33}  & \textbf{12.17}   & \textbf{11.17}   & \textbf{8.00}    & \textbf{10.67}   & \textbf{18.90}   & \textbf{19.43}   & \textbf{12.57}   & \textbf{16.80}   & \textbf{19.33}   & \textbf{9.80}    & \textbf{9.37}    & \textbf{11.95}           \\
Relative WER (\%)$\uparrow$                                                   & \textbf{38.75\%} & \textbf{42.75\%} & \textbf{19.79\%} & \textbf{45.42\%} & \textbf{25.60\%} & \textbf{33.16\%} & \textbf{7.41\%} & \textbf{36.63\%} & \textbf{23.52\%} & \textbf{23.81\%} & \textbf{30.28\%} & \textbf{23.79\%} & \textbf{12.86\%} & \textbf{19.96\%} & \textbf{36.12\%} & \textbf{27.04\%} & \textbf{42.69\%} & \textbf{27.39\%} & \textbf{28.73\%}         \\ \bottomrule
% Instruction Finetuning                                            & \textbf{4.70}    & \textbf{8.17}    & \textbf{11.43}   & \textbf{11.90}   & \textbf{9.63}    & \textbf{14.43}   & 14.80   & \textbf{15.23}   & \textbf{11.30}   & \textbf{9.73}   & \textbf{12.83}   & \textbf{21.13}   & \textbf{20.90}  & \textbf{13.90}   & \textbf{19.43}   & \textbf{20.97}   & \textbf{10.73}   & \textbf{10.70}   & \textbf{13.44}           \\
% %WER Relative \%  $\uparrow$                                                 & \textbf{41.25\%} & \textbf{37.66\%} & \textbf{10.68\%} & \textbf{34.62\%} & \textbf{13.99\%} & \textbf{24.04\%} & -2.78\% & \textbf{20.66\%} & \textbf{22.60\%} & \textbf{7.30\%} & \textbf{16.12\%} & \textbf{14.78\%} & \textbf{6.28\%} & \textbf{11.46\%} & \textbf{26.11\%} & \textbf{20.88\%} & \textbf{37.23\%} & \textbf{17.05\%} & \textbf{19.86\%}         \\ \bottomrule
% Relative WER Reduction\%  $\uparrow$                                                 & \textbf{41.25} & \textbf{37.66} & \textbf{10.68} & \textbf{34.62} & \textbf{13.99} & \textbf{24.04} & -2.78 & \textbf{20.66} & \textbf{22.60} & \textbf{7.30} & \textbf{16.12} & \textbf{14.78} & \textbf{6.28} & \textbf{11.46} & \textbf{26.11} & \textbf{20.88} & \textbf{37.23} & \textbf{17.05} & \textbf{19.86}         \\ \bottomrule
\end{tabular}
\end{adjustbox}
\caption{\textbf{ASR Adaptation with Synthetic Text Corpus.} 
%To adapt source domain ASR models to new target domains with \emph{no real text or speech data} from the target domain, we prepare a synthetic text corpus using LLMs with ICIF (Section \ref{sec:instuct}) and a corresponding synthetic speech corpus using CSS (Section \ref{sec:css}). 
Results of ASR models finetuned on target domain synthetic data from our pipeline with ICIF. For each target domain, the source domain ASR (baseline) is trained on LibriSpeech followed by the data from 17 domains (excluding the target domain) in SLURP dataset.
The metric shown is WER (lower is better). } 
\label{tab:11000table}
\end{table*}
\vspace{-0.1in}
\subsection{Controllable Speech Synthesis}
\label{sec:css}
%After LLMs prepare a decent target domain text data, we employ a Controllable Speech Synthesis (CSS) Model to synthesize speech given text corpus $y^t_i \sim p_{CSS}(y|x_i)$.

We use a state-of-the-art Controllable Speech Synthesis (CSS) model \cite{chang2022style} to
synthesize speech $y^t_i \sim p_{CSS}(y|x^t_i)$, given target domain text $x^t_i$ generated by the instruction finetuned LLM model. 
%Different from the mainstream text-to-speech (TTS) models %\cite{shen2018natural,ren2020fastspeech},
The CSS model %\cite{chang2022style} 
%To increase the acoustic diversity, we adopt Style Equalization \cite{chang2022style}, which is designed based on Variational Recurrent Neural Net (VRNN) \cite{chung2015recurrent} and 
learns a prior distribution to model the acoustic style of speech. By sampling from this prior distribution, the model can produce a synthetic speech corpus in various acoustic conditions.
%We use CSS model to synthesize various styles of speech $y^t_i \sim p_{CSS}(y|x_i)$ for downstream ASR finetuning.
% For CSS model, we adopt Style Equalization \cite{chang2022style}, which is designed based on Variational Recurrent Neural Net (VRNN) \cite{chung2015recurrent} and learns a prior distribution to model the acoustic style of speech.
\vspace{-0.1in}
\subsection{ASR Model Adaptation}
Finally, we finetune the ASR model on the synthetic speech prepared by the LLM and CSS model. In our initial experiments, we observed that the ASR model usually overfits to synthetic speech artifacts during finetuning, which limits its performance. 
%To address this problem, we add a small portion of the source domain real speech data along with the target domain synthesized speech data to regularize the ASR model's performance. 
To address this problem, we add real speech data (\emph{i.e.,} from source domains) to the synthetic speech from the target domain to regularize the ASR model finetuning. 
%

%
% To further enhance this capability, we make the following three simple modifications:
% %
% (1) increasing the number of gaussian mixtures of VRNN output distribution (from 3 to 10)
% (2) enlarge the size of acoustic style feature (from 512 to 768)
% (3) initialize the hidden states of VRNN using the average of the style vector sequence.
% %
% In this work, we train the modified Style Equalization model on the training set of LibriTTS \cite{zen2019libritts}.

\section{Experimental Setup}
\label{sec:experiments}
\subsection{Dataset}
%SLURP dataset \cite{bastianellietal2020slurp} is a spoken language understanding (SLU) dataset which consists of human's commands towards the robot, using 200 pre-defined prompts such as “How would you ask for the time.
%We split SLURP data according to 
SLURP \cite{bastianellietal2020slurp} is a spoken language understanding dataset containing 16521 utterances of human commands towards a virtual agent, based on 200 pre-defined prompts such as ``How would you ask for the time.'' 
The utterances are recorded in two types of acoustic environments (headset and far-field), and categorized into 18 domains (email, alarm, and takeaway, etc.). 
%[specify headset only]
%In this work, we use the pre-defined 18 domains to form the experimental setup of target data free ASR domain adaptation.
%
%Specifically, we combine 17 domains in SLURP to form the source domain, and use the remaining one as the target domain in testing phase.
We use the headset subset to conduct experiments of zero-shot ASR domain adaptation.
In each of our experiments, we select one of these domains as the target domain %for zero-shot adaptation 
and combine the remaining 17 domains to form the source domain.
Our goal is to improve the performance of a pretrained source domain ASR model on the target domain, without using any real speech or real text data from the target domain.
%
%By conducting 18-fold cross validation, w
%We report the word error rate (WER) of each finetuned ASR model on the official test split of its respective target domain.
\subsection{Large Language Models} 
We use LLaMA-7B \cite{touvron2023llama} to synthesize the text corpus. LLaMA is a state-of-the-art LLM with a decoder-based transformer architecture that is pretrained on trillions of tokens. LLaMA has shown % good ability on various benchmarks and
excellent performance on downstream tasks with instruction finetuning \cite{touvron2023llama, 10.1162/tacl_a_00536}.
%\todo {need citation}
We apply low-rank adaptation (LoRA) \cite{hu2021lora} to freeze most of the model parameters and improve efficiency of instruction finetuning. During inference/synthesis, we follow \cite{lin-etal-2023-selective} to use typical decoding \cite{10.1162/tacl_a_00536} with $\tau = 0.9$ and set the repetition penalty \cite{keskar2019ctrl} to $1.1$. We include $10$ demonstrations in the inference prompt.
\subsection{Controllable Speech Synthesis (CSS) }
Our CSS model is adopted from Style Equalization \cite{chang2022style}, which is based on a Variational Recurrent Neural Net (VRNN).
We make the following four modifications to enhance its acoustic style modeling:
(1) increasing the number of Gaussian mixtures of VRNN output distribution (from 3 to 10); 
(2) increasing the size of acoustic style feature (from 512 to 768);
(3) initializing the hidden states of VRNN using the average of the style vector sequence; and
(4) using the acoustic style feature to modulate the output linear layers, similar to what is done in \cite{chan2021pi}.
We train the modified CSS model on the training set of LibriTTS \cite{zen2019libritts}.
\subsection{ASR Model Adaptation}
We use ESPNet \cite{watanabe2018espnet} to build the E2E ASR model, which is composed of a conformer-based encoder \cite{gulati2020conformer} and a transformer-based decoder \cite{vaswani2017attention}.
%
%In our experiment, we treat one domain in SLURP as the target domain without any speech and text data, and the combination of the remaining 17 domains as the source domain real dataset. 
%
In each of our experiments, we first obtain a pretrained source domain ASR model by training on LibriSpeech \cite{panayotov2015librispeech} followed by the source domain data (\emph{i.e.,} 17 pre-defined SLURP domains excluding the target domain).
%To obtain the source domain ASR model, we first pre-train the model on LibriSpeech \cite{panayotov2015librispeech}, and then finetune it on the source domain real dataset.
%
%In addition to target domain synthetic speech, we also mix with $10 \%$ source domain real speech to finetune ASR models.
%
We then adapt this pretrained ASR model to the target domain using the synthetic data from LLM and CSS. For fair comparison between models, we select all final checkpoints using the target domain development set as a validation set.
%The final checkpoints of both the source domain ASR model and the target domain finetuned ASR model are selected according to their performance on target domain development set. 
%\subsection{Baselines}
%In addition to instruction finetuning, we consider two common methods as comparable baselines for LLMs to synthesize the text corpus: prompting pretrained LLMs directly with designed prompt and in-context learning LLM.
%\subsubsection{Prompt Directly (PD) }
%Recent works have shown that pretrained LLMs have good ability to complete a sentence by providing only text in zero-shot fashion.
%To adjust PD to our scenario, we directly prompt pretrained LLaMA using the text  `\textit{Please generate a sentence related to $d_t$:}', without providing any demonstrations from source data or finetuning the LLaMA model.

%\vspace{-0.1in}
\section{Results and Discussion}
%\vspace{-0.1in}
% \begin{table}[!]
% \begin{adjustbox}{width=0.5\textwidth}

% \begin{tabular}{cccc}
% \toprule
% Word Error Rate (\%)                                                             & \multicolumn{3}{c}{Target Domains}                                        \\ \hline
% Methods                                                                     & Email                    & Alarm                   & Takeaway              \\ \midrule
% \begin{tabular}[c]{@{}c@{}}Original ASR\\ (Baseline)\end{tabular}           & 19.0                     & 8.0                     & 26.5                  \\ \midrule
% Prompt directly                                                             & 18.1 (7\%)               & 7.433 (4.7\%)           & 20.7 (21.9 \%)        \\ \midrule
% \multicolumn{1}{l}{In-Context Learning}                                     & \multicolumn{1}{l}{16.53 (12.98\%)}     & \multicolumn{1}{l}{5.77 (27.92 \%)}    & \multicolumn{1}{l}{22.07 (16.73\%)}  \\
% Instruction Finetuning                                                      & \textbf{15.43 (18.79\%)} & \textbf{4.43 (42.9 \%)} & \textbf{17.77 (33\%)} \\ \midrule
% \begin{tabular}[c]{@{}c@{}}Real Text Only\\ (Access Real Text)\end{tabular} & 12.85                    & 5.375                   & 14.475                \\ \bottomrule
% \end{tabular}
% \end{adjustbox}
% \caption{\textbf{ASR performance with different synthetic text data.} The LLMs with instruction finetuned synthesize the text corpus to finetune ASR models. The metrics we use is WER (lower is better). } 
%     \label{tab:11000table}
% \end{table}
\subsection{ASR Adaptation with Synthetic Text Corpus}
In Table~\ref{tab:11000table}, we report the performance of ASR models finetuned on target domain data from our corpus synthesis pipeline.
%We evaluate the ASR models finetuned on the target domain data synthesized by our corpus synthesis pipeline.
For each target domain, we prepare the synthetic text corpus using LLMs with %LLaMA with in-context instruction finetuning (ICIF) 
ICIF and the corresponding synthetic speech using CSS. 
%We report the word error rate (WER) on the testing splits of 18 domains in SLURP dataset.
%The results in Table \ref{tab:11000table} indicate that the ASR models finetuned on the corpus prepared by our proposed method gain large improvements (average 28.73\%).
Remarkably, we achieve large reductions in WER across the board (average relative improvement of $28.73\%$), without using any real text or speech from the target domain for finetuning.
%The results in Table \ref{tab:11000table} indicate that the ASR models finetuned on the corpus prepared by our proposed method gain large improvements (average 28.73\%).
For some target domains (\emph{i.e.,} \textit{Audio}, \textit{Cooking}, and \textit{Transport}), we achieve more than $40\%$ relative improvement compared to the pretrained source domain models. 
In addition, the finetuned ASR models also yield a small improvement (average relative WER reduction of $5.98\%$) in source domains. 
Overall, these results demonstrate the efficacy of our corpus synthesis pipeline for adapting ASR models to unseen text domains.

%An oracle experiment shows that the average WER is further reduced (from 11.95\% to 10.77\%) with real target domain text and synthetic speech from CSS. 
%It indicates the potential room for improvement of our method.

We also finetune the source domain ASR models with (1) real target domain text and synthetic speech, and (2) real text and real speech, receiving average WERs of 10.77\% and 10.74\%, respectively. 
Note that, the purpose of this experiment is to establish an upper-bound for adaptation, and real target domain data is not available in zero-shot adaptation.
\vspace{-0.05in}
\subsection{Analysis of ICIF}
\label{sec:size}
\vspace{-0.05in}
%%% Will need to revisit QQ
\textbf{Contribution of IF and ICL}

As detailed in Section~\ref{sec:instuct}, ICIF involves two steps: (1) \emph{instruction} (IF), which finetunes the LLM using instructions formulated from a source text corpus, and (2) \emph{demonstration} (ICL), which prompts the LLM with some example instructions. Table~\ref{tab:method_compared} analyzes the individual contributions of these components.  %study the impact of instruction and demonstration by isolating out each of method independently, and the results are in Table \ref{tab:method_compared}.
We observe that both are useful for improving the WER of the finetuned ASR model: using either instruction (IF) or demonstration (ICL) improves the WER over naive prompting (\emph{i.e.,} from $14.02$ to $12.13$ and $12.59$ respectively). Combining IF and ICL (ICIF) further improves the WER to $11.95$. 
%If we took off the instruction finetuning ($-$Instruct.), WER increase from $11.95$ to $12.13$. And if we didn't perform any demonstrations ($-$Demo.), WER increase to $12.59$. Furthermore, if we took off both which is only using a prompt \textit{"please generate a sentence related to $d_t$:"} to prompt pretrained LLaMA. WER increase significantly to $16.40$. 
These results indicate that both instruction and demonstration are useful to our synthetic corpus pipeline. 
Next, we ask whether instruction and demonstration have overlapping effects on the synthetic text quality, or whether they play distinct roles. To address this question, we profile the synthetic text along two additional axes: (1) diversity, measured by Self-BLEU 4-gram \cite{zhu2018texygen} and (2) similarity to the real target corpus, measured by the JS divergence between token distributions \cite{ruder2017data}. As shown in Table~\ref{tab:method_compared}, instruction (IF) is highly effective at generating text similar to the target domain, but at the cost of diversity. On the other hand, demonstration (ICL) achieves high diversity with a modest improvement in similarity. Combining the two techniques strikes a balance between improving diversity and similarity of the synthetic text to the target domain. We conclude that ICIF enables the LLM to map from domain names to more relevant and diverse texts,  which in turn improves the generalization of ASR models to unseen target domains.
\begin{table}[t!]
\begin{adjustbox}{width=\linewidth}
\begin{tabular}{ccccc}
\toprule
 & WER $\downarrow$ & \begin{tabular}[c]{@{}c@{}} Relative WER \\ Improvement (\%) $\uparrow$\end{tabular} & \begin{tabular}[c]{@{}c@{}}Diversity $\downarrow$\\ (SB-4)\end{tabular} & \begin{tabular}[c]{@{}c@{}}Similarity $\downarrow$\\ (JS-Div)\end{tabular} \\ \midrule
Source Domain ASR & 16.77 & - & - & - \\ \midrule
ICIF (IF+ICL) & \textbf{11.95} & \textbf{28.73} & 0.596 & 0.466 \\
Demo (ICL) & 12.13 & 27.67 & \textbf{0.424} & 0.482 \\
Instruct (IF) & 12.59 & 24.92 & 0.74 & \textbf{0.451} \\
Naive Prompting & 14.02 & 16.40 & 0.471 & 0.521 \\ 
% \midrule
% Real Text (Oracle) & 10.77 & - & - & - \\
% Real Speech (Oracle) & 10.74 & - & - & - \\ 
\bottomrule
\end{tabular}
\end{adjustbox}
\caption{\textbf{Analysis of ICIF.} We investigate the individual contributions of instruction (IF) and demonstration (ICL) to ICIF. %We also report the ASR finetuned with target domain real speech (Real Speech) and synthetic speech with target domain real text (Real Text). 
In addition to the WER, we report the diversity of the synthetic text (SB-4), and its similarity to the target domain text corpus (JS-Div).
See Section~\ref{sec:size} for details. }
\label{tab:method_compared}
\end{table}

\noindent\textbf{Impact of synthetic text corpus size on WER}

\begin{figure}[t!]
    \begin{adjustbox}{width=\linewidth}
    \centering
    \includegraphics{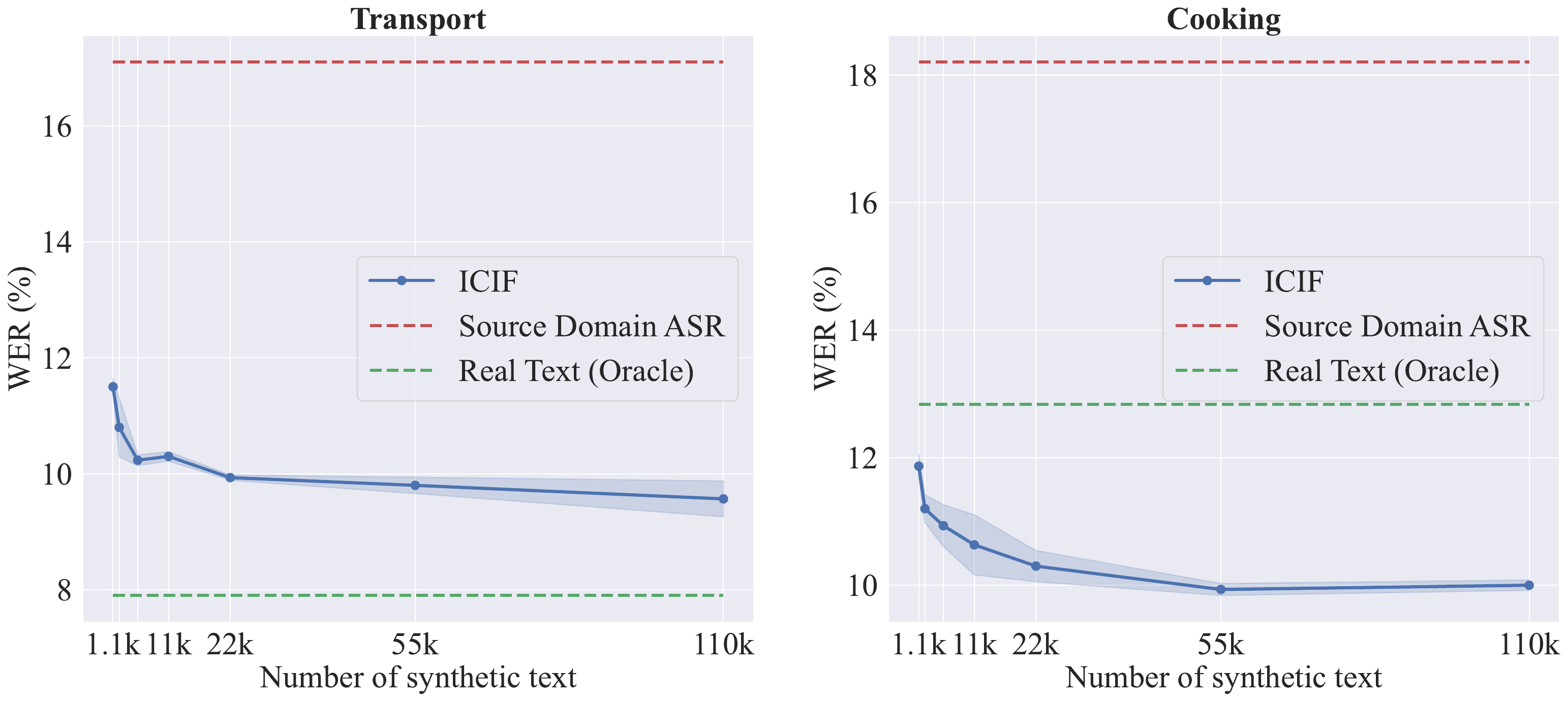}
    \end{adjustbox}
    \caption{\textbf{Number of synthetic text v. WER.} We vary the number of synthetic text samples used to finetune the ASR models and report the WER for two randomly-selected target domains. Number of samples and WER are shown on the x and y axes respectively.}
    % \caption{\textbf{Number of synthetic text versus WER.}}
    \label{fig:tansport}
\end{figure}

%Fig.~\ref{fig:tansport} shows the experimental results from using various amount of synthetic text to finetune ASR models, and its corresponding WER. 
Fig.~\ref{fig:tansport} shows the performance of ASR models finetuned on varying amounts of synthetic text for two randomly-selected target domains (\textit{`Transport'} and \textit{`Cooking'}). 
%% Ting-Yao: need to be revisited
%In the figure, the $x$ axis is the number of synthetic text used to finetune ASR model and the $y$ axis is the absolute WER. We plot the figures on \textit{`Transport'} and \textit{`Cooking'} domains.
In general, we find that using more synthetic text data to finetune the ASR models improves the WER, which suggests that the models benefit from exposure to greater text diversity. On the other hand, we also observe that ASR performance saturates at some point (\emph{e.g.,} at around 55K samples for the \emph{``Cooking''} domain). This may be due to synthetic artifacts or noise. We leave the problem of synthetic data selection to future work.
%In the figure, we can observe that when we use more synthetic data to finetune ASR model, the increased text diversity make the performance keeps going up. On the other side, we can also observe ASR performance saturated on 55k synthetic data, we believe it's because there are some synthetic artifact as noise that decrease the performance. We leave this as a future work to eliminate outliers from synthetic data.

\noindent\textbf{Impact of number of demonstrations on WER}

\begin{figure}[t!]
    \begin{adjustbox}{width=\linewidth}
    \centering
    \includegraphics{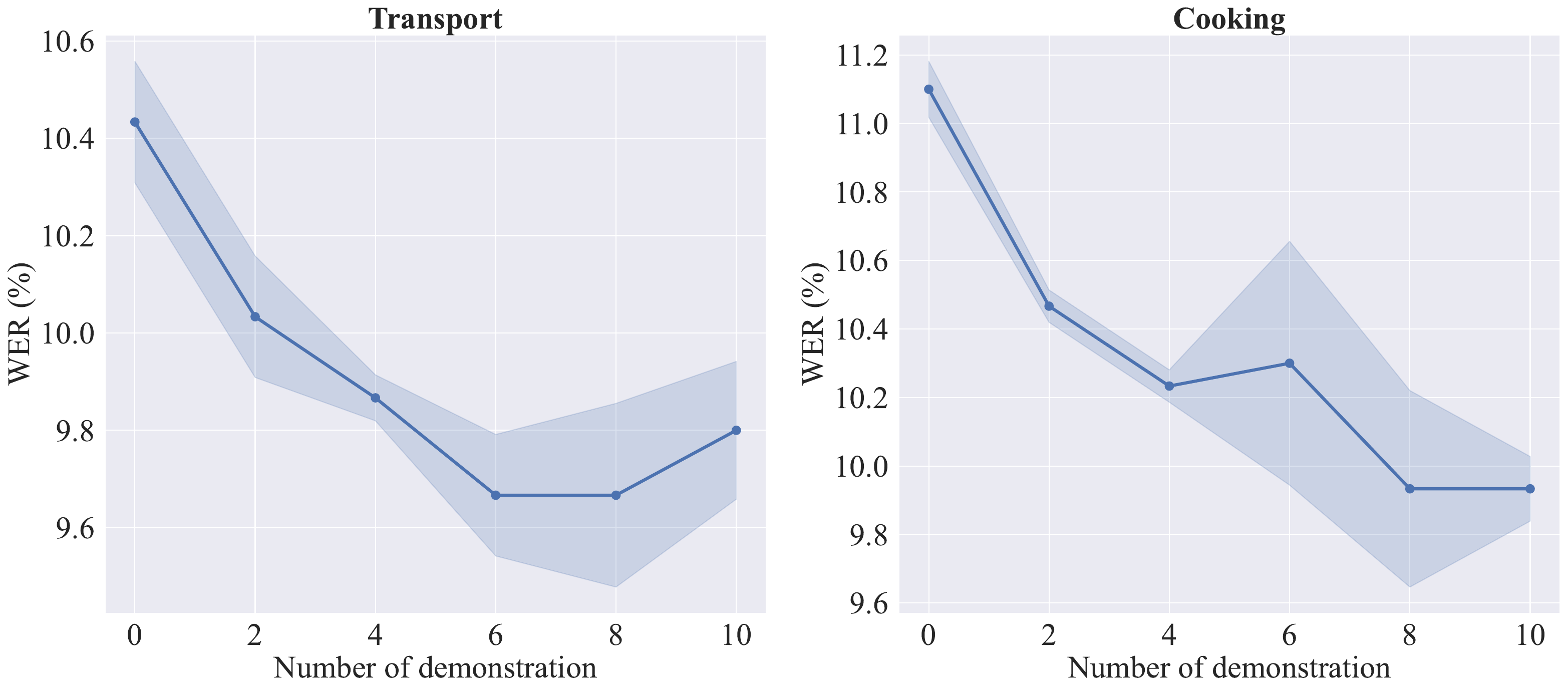}
    \end{adjustbox}
    \caption{\textbf{Number of demonstrations v. WER.} We vary the number of demonstrations used for promting the LLM model and report the WER of finetuned ASR models for two randomly-selected target domains. Number of demonstrations and WER are shown on the x and y axes respectively.}
    % \caption{\textbf{Number of synthetic text versus WER.}}
    \label{fig:num_dmo}
\end{figure}

%\subsubsection{Does more demonstrations decrease WER?}
Since demonstrations increase synthetic text diversity, we also investigate the impact of the number of demonstrations on the performance of finetuned ASR models. Fig.~\ref{fig:num_dmo} shows the WER on two randomly-selected target domains when varying the number of demonstrations from $0$ to $10$. We observe that WER is improved significantly even with two demonstrations and continues to improve with more demonstrations. Interestingly, we also observe that the standard deviation of the WER increases with more demonstrations. We hypothesize this is due to increased text diversity, which leads to variable outcomes during finetuning. The selection and ordering of demonstrations may also impact the synthetic text quality. We leave these investigations to future work.
%In this section, we study the influence of different number of demonstration used in ICIF towards WER. The results are included in Fig.~\ref{fig:num_dmo}, we can observe that WER decrease significantly even we demonstrates two examples to LLaMA. 
%Furthermore, WER keeps decreasing when we increase the number of demonstrations, and this shows that more observations can help ASR generalize to new domains. However, we also observe that standard deviation become larger with more demonstrations. This implies (1) the synthetic texts are more diverse (2) the demonstrations order/selections impact ASR performance. We also leave this as future work. 

\vspace{-0.05in}
\section{Conclusions}
\vspace{-0.05in}
In this paper, we propose a pipeline which consists of a LLM and a CSS model to adapt ASR models with synthesize speech corpus.
We apply the data synthesis pipeline to ASR domain adaptation with no target domain data, and receive $16 \%$ relative improvements with pretrained LLMs.
To further improve synthesized text quality, we employ an innovative in-context instruction finetuning (ICIF) method on LLMs.
% To further improve synthesized text quality, we investigate two strategies, instruction finetuning (IF) and In-context Learning (ICL).
%The results shows that we can get $28 \%$ WER relative improvements on average with our proposed method.
The results show that our proposed method yields $28 \%$ average WER relative improvement on unseen target domains without dropping the performance on source domains.
% In this paper, we aim at adapting ASR models to new unseen domains without accessing any text or speech data from that domains.
% To solve this problem, we propose a novel data synthesis pipeline that consists of a LLM and a CSS models to prepare speech corpus to finetune ASR models.
% Furthermore, we propose a new method 

{\small
\bibliographystyle{IEEEbib}
\bibliography{strings,refs}

\begin{thebibliography}{10}

\bibitem{meng2021internal}
Zhong Meng, Yashesh Gaur, Naoyuki Kanda, Jinyu Li, Xie Chen, Yu~Wu, and Yifan
  Gong,
\newblock ``Internal language model adaptation with text-only data for
  end-to-end speech recognition,''
\newblock {\em Proc. InterSpeech}, 2022.

\bibitem{pylkkonen2021fast}
Janne Pylkk{\"o}nen, Antti Ukkonen, Juho Kilpikoski, Samu Tamminen, and Hannes
  Heikinheimo,
\newblock ``Fast text-only domain adaptation of rnn-transducer prediction
  network,''
\newblock {\em Proc. InterSpeech}, 2021.

\bibitem{deng2023adaptable}
Keqi Deng and Philip~C Woodland,
\newblock ``Adaptable end-to-end asr models using replaceable internal lms and
  residual softmax,''
\newblock in {\em ICASSP}. IEEE, 2023, pp. 1--5.

\bibitem{mittal2022situ}
Ashish Mittal, Sunita Sarawagi, and Preethi Jyothi,
\newblock ``In-situ text-only adaptation of speech models with low-overhead
  speech imputations,''
\newblock in {\em ICLR}, 2022.

\bibitem{joshi2022simple}
Raviraj Joshi and Anupam Singh,
\newblock ``A simple baseline for domain adaptation in end to end asr systems
  using synthetic data,''
\newblock {\em arXiv preprint arXiv:2206.13240}, 2022.

\bibitem{bataev2023text}
Vladimir Bataev, Roman Korostik, Evgeny Shabalin, Vitaly Lavrukhin, and Boris
  Ginsburg,
\newblock ``Text-only domain adaptation for end-to-end asr using integrated
  text-to-mel-spectrogram generator,''
\newblock {\em Proc. InterSpeech}, 2023.

\bibitem{Dingliwa2022}
Saket Dingliwa, Ashish Shenoy, Sravan Bodapati, Ankur Gandhe, Ravi~Teja Gadde,
  and Katrin Kirchhoff,
\newblock ``Domain prompts: Towards memory and compute efficient domain
  adaptation of asr systems,''
\newblock in {\em Proc. Interspeech}, 2022.

\bibitem{ma2023can}
Rao Ma, Mengjie Qian, Potsawee Manakul, Mark Gales, and Kate Knill,
\newblock ``Can generative large language models perform asr error
  correction?,''
\newblock {\em arXiv preprint arXiv:2307.04172}, 2023.

\bibitem{li2023prompting}
Yuang Li, Yu~Wu, Jinyu Li, and Shujie Liu,
\newblock ``Prompting large language models for zero-shot domain adaptation in
  speech recognition,''
\newblock {\em arXiv preprint arXiv:2306.16007}, 2023.

\bibitem{chang2022style}
Jen-Hao~Rick Chang, Ashish Shrivastava, Hema Koppula, Xiaoshuai Zhang, and
  Oncel Tuzel,
\newblock ``Style equalization: Unsupervised learning of controllable
  generative sequence models,''
\newblock in {\em Proc. ICML}, 2022.

\bibitem{bastianellietal2020slurp}
Emanuele Bastianelli, Andrea Vanzo, Pawel Swietojanski, and Verena Rieser,
\newblock ``{SLURP}: A spoken language understanding resource package,''
\newblock in {\em Proc. EMNLP}, 2020.

\bibitem{ye-etal-2022-zerogen}
Jiacheng Ye, Jiahui Gao, Qintong Li, Hang Xu, Jiangtao Feng, Zhiyong Wu, Tao
  Yu, and Lingpeng Kong,
\newblock ``{Z}ero{G}en: Efficient zero-shot learning via dataset generation,''
\newblock in {\em Proc. EMNLP}, 2022.

\bibitem{yoo-etal-2021-gpt3mix-leveraging}
Kang~Min Yoo, Dongju Park, Jaewook Kang, Sang-Woo Lee, and Woomyoung Park,
\newblock ``{GPT}3{M}ix: Leveraging large-scale language models for text
  augmentation,''
\newblock in {\em Findings of the Association for Computational Linguistics:
  EMNLP 2021}, 2021.

\bibitem{meng2022generating}
Yu~Meng, Jiaxin Huang, Yu~Zhang, and Jiawei Han,
\newblock ``Generating training data with language models: Towards zero-shot
  language understanding,''
\newblock in {\em NeurIPS}, 2022.

\bibitem{chung2022scaling}
Hyung~Won Chung, Le~Hou, Shayne Longpre, Barret Zoph, Yi~Tay, William Fedus,
  Eric Li, Xuezhi Wang, Mostafa Dehghani, Siddhartha Brahma, et~al.,
\newblock ``Scaling instruction-finetuned language models,''
\newblock {\em arXiv preprint arXiv:2210.11416}, 2022.

\bibitem{brown2020language}
Tom Brown, Benjamin Mann, Nick Ryder, Melanie Subbiah, Jared~D Kaplan, Prafulla
  Dhariwal, Arvind Neelakantan, Pranav Shyam, Girish Sastry, Amanda Askell,
  et~al.,
\newblock ``Language models are few-shot learners,''
\newblock {\em NeurIPS}, 2020.

\bibitem{touvron2023llama}
Hugo Touvron, Thibaut Lavril, Gautier Izacard, Xavier Martinet, Marie-Anne
  Lachaux, Timothée Lacroix, Baptiste Rozière, Naman Goyal, Eric Hambro,
  Faisal Azhar, Aurelien Rodriguez, Armand Joulin, Edouard Grave, and Guillaume
  Lample,
\newblock ``Llama: Open and efficient foundation language models,'' 2023.

\bibitem{10.1162/tacl_a_00536}
Clara Meister, Tiago Pimentel, Gian Wiher, and Ryan Cotterell,
\newblock ``Locally typical sampling,''
\newblock {\em Trans. of ACL}, vol. 11, pp. 102--121, 2023.

\bibitem{hu2021lora}
Edward~J Hu, Yelong Shen, Phillip Wallis, Zeyuan Allen-Zhu, Yuanzhi Li, Shean
  Wang, Lu~Wang, and Weizhu Chen,
\newblock ``Lora: Low-rank adaptation of large language models,''
\newblock {\em Proc. ICLR}, 2022.

\bibitem{lin-etal-2023-selective}
Yen-Ting Lin, Alexandros Papangelis, Seokhwan Kim, Sungjin Lee, Devamanyu
  Hazarika, Mahdi Namazifar, Di~Jin, Yang Liu, and Dilek Hakkani-Tur,
\newblock ``Selective in-context data augmentation for intent detection using
  pointwise {V}-information,''
\newblock in {\em Proc. EACL}, 2023.

\bibitem{keskar2019ctrl}
Nitish~Shirish Keskar, Bryan McCann, Lav~R Varshney, Caiming Xiong, and Richard
  Socher,
\newblock ``Ctrl: A conditional transformer language model for controllable
  generation,''
\newblock {\em arXiv preprint arXiv:1909.05858}, 2019.

\bibitem{chan2021pi}
Eric~R Chan, Marco Monteiro, Petr Kellnhofer, Jiajun Wu, and Gordon Wetzstein,
\newblock ``pi-gan: Periodic implicit generative adversarial networks for
  3d-aware image synthesis,''
\newblock in {\em Proc. CVPR}, 2021.

\bibitem{zen2019libritts}
Heiga Zen, Viet Dang, Rob Clark, Yu~Zhang, Ron~J Weiss, Ye~Jia, Zhifeng Chen,
  and Yonghui Wu,
\newblock ``Libritts: A corpus derived from librispeech for text-to-speech,''
\newblock {\em arXiv preprint arXiv:1904.02882}, 2019.

\bibitem{watanabe2018espnet}
Shinji Watanabe, Takaaki Hori, Shigeki Karita, Tomoki Hayashi, Jiro Nishitoba,
  Yuya Unno, Nelson Enrique~Yalta Soplin, Jahn Heymann, Matthew Wiesner, Nanxin
  Chen, et~al.,
\newblock ``Espnet: End-to-end speech processing toolkit,''
\newblock {\em arXiv preprint arXiv:1804.00015}, 2018.

\bibitem{gulati2020conformer}
Anmol Gulati, James Qin, Chung-Cheng Chiu, Niki Parmar, Yu~Zhang, Jiahui Yu,
  Wei Han, Shibo Wang, Zhengdong Zhang, Yonghui Wu, et~al.,
\newblock ``Conformer: Convolution-augmented transformer for speech
  recognition,''
\newblock {\em Proc. InterSpeech}, 2020.

\bibitem{vaswani2017attention}
Ashish Vaswani, Noam Shazeer, Niki Parmar, Jakob Uszkoreit, Llion Jones,
  Aidan~N Gomez, {\L}ukasz Kaiser, and Illia Polosukhin,
\newblock ``Attention is all you need,''
\newblock {\em NeurIPS}, 2017.

\bibitem{panayotov2015librispeech}
Vassil Panayotov, Guoguo Chen, Daniel Povey, and Sanjeev Khudanpur,
\newblock ``Librispeech: an asr corpus based on public domain audio books,''
\newblock in {\em Proc. ICASSP}, 2015.

\bibitem{zhu2018texygen}
Yaoming Zhu, Sidi Lu, Lei Zheng, Jiaxian Guo, Weinan Zhang, Jun Wang, and Yong
  Yu,
\newblock ``Texygen: A benchmarking platform for text generation models,''
\newblock {\em SIGIR}, 2018.

\bibitem{ruder2017data}
Sebastian Ruder, Parsa Ghaffari, and John~G. Breslin,
\newblock ``Data selection strategies for multi-domain sentiment analysis,''
  2017.

\end{thebibliography}
}
\end{document}